\documentclass[12pt]{article}

\usepackage{graphicx}

\begin{document}


\title{\bf
MONOPOLE-ANTIMONOPOLE PAIR DYONS} 

\author{
{\bf Kok-Geng Lim, Rosy Teh and Khai-Ming Wong}\\
{\normalsize School of Physics, Universiti Sains Malaysia}\\
{\normalsize 11800 USM Penang, Malaysia}}

\date{July, 2010}
\maketitle

\begin{abstract}
Monopole-antimonopole pair (MAP) with both electric and magnetic charges are presented. The MAP possess opposite magnetic charges but they carry the same electric charges. These stationary MAP dyon solutions possess finite energy but they do not satisfy the first order Bogomol'nyi equations and are not BPS solutions. They are axially symmetric solutions and are characterized by a parameter, $-1\leq\eta\leq 1$ which determines the net electric charges of these MAP dyons. These dyon solutions are solved numerically when the magnetic charges of the dipoles are $n=\pm 1, \pm 2$ and when the strength of the Higgs field potential $\lambda=0, 1$. When $\lambda=0$, the time component of the gauge field potential is parallel to the Higgs field in isospin space and the MAP separation distance, total energy and net electric charge increase exponentially fast to infinity when $\eta$ approaches $\pm 1$. However when $\lambda=1$, all these three quantities approach a finite critical value as $\eta$ approaches $\pm 1$. 
\end{abstract}


\section{Introduction}

The SU(2) Yang-Mills-Higgs (YMH) field theory in $3+1$ dimensions, with the Higgs field in the adjoint representation possess magnetic monopoles solutions \cite{kn:1}-\cite{kn:2} as well as dyons solutions \cite{kn:3}. The 't Hooft-Polyakov monopole solution with non zero Higgs mass and self-interaction is the first monopole solution that possess finite energy and this solution can acquire an electric charge to become a dyon \cite{kn:3}. These numerical, spherically symmetric monopole and dyon solutions of unit magnetic charge are invariant under a U(1) subgroup of the local SU(2) gauge group. Exact monopoles and multimonopoles solutions exist only in the Bogomol'nyi-Prasad-Sommerfield (BPS) limit \cite{kn:2}. Outside of this limit, when the Higgs field potential is non vanishing only numerical solutions exist. The exact dyon solutions found by Prasad and Sommerfield \cite{kn:3} in the BPS limit is stable as they are the absolute minima of the energy \cite{kn:4}.

The word,``dyon" was coined by J. Schwinger in 1969 \cite{kn:5} for a particle that possesses both magnetic and electrical charges. A dyon with a fixed magnetic charge can possesses varying electric charges. The dyons solutions of Julia and Zee \cite{kn:3} are time independent solutions that possess non vanishing kinetic energy. Dyons with axial symmetry were constructed by B. Hartmann et al. \cite{kn:6} when $0\leq\eta\leq1$. These axial dyons are actually 't-Hooft-Polyakov monopoles that possess magnetic charges, $n=1, 2, 3$ and non vanishing electric charges when the parameter $\eta$ is nonzero. It was found that when the strength of the Higgs potential $\lambda$ is non vanishing, the total energy and total electric charge of the system approach finite critical values when the parameter $\eta$ approaches 1. However when $\lambda=0$,  the total energy and total electric charge of the system approach infinity when the parameter $\eta$ approaches 1. We found that the MAP dyon possesses similar characteristics. 

In this paper, we would like to present the axially symmetric MAP dyons by introducing electric charge to the system. The procedure of introducing electric charges to the monopole solution is standard as was first shown by Julia and Zee in 1975 \cite{kn:3}. The monopole and antimonopole carry the same electric charges and hence experience a repulsion force due to the electric charge. These MAP dyon solutions possess finite energy but they do not satisfy the first order Bogomol'nyi equations and are not BPS solutions. They are characterized by a parameter, $-1\leq\eta\leq 1$ which determines the net electric charges of these MAP dyons. When $\eta$ is positive (negative), the electric charges carried by both monopole and antimonopole are positive (negative). These dyons solutions are solved numerically when the magnetic charges of the dipoles are $n=\pm 1, \pm 2$ and when the strength of the Higgs field potential $\lambda=0, 1$. When $\lambda=0$, the time component of the gauge field potential is parallel to the Higgs field in isospin space and the MAP separation distance, total energy and net electric charge of the MAP dyons increase exponentially fast to infinity when $\eta$ approaches $\pm 1$. However when $\lambda=1$, all these three quantities approach a critical finite value when $\eta$ approaches $\pm 1$. 

We briefly review the SU(2) Yang-Mills-Higgs field theory and the ansatz in the next section. We present our MAP dyon solutions and the results of our numerical calculations in section 3. We end with some comments in section 4.

\section{The SU(2) Yang-Mills-Higgs Theory}
\label{sect:2}
The SU(2) YMH Lagrangian in 3+1 dimensions is given by
\begin{equation}
{\cal L} = -\frac{1}{4}F^a_{\mu\nu} F^{a\mu\nu} - \frac{1}{2}D^\mu \Phi^a D_\mu \Phi^a - \frac{1}{4}\lambda(\Phi^a\Phi^a - \frac{\mu^2}{\lambda})^2, 
\label{eq.1}
\end{equation}

\noindent where $\mu$ is the Higgs field mass, $\lambda$ is the strength of the Higgs potential and $\xi = \mu/\sqrt{\lambda}$ is the vacuum expectation value of the Higgs field. The Lagrangian (\ref{eq.1}) is gauge invariant under the set of independent local SU(2) transformations at each space-time point.
The covariant derivative of the Higgs field and the gauge field strength tensor are given respectively by 
\begin{eqnarray}
D_{\mu}\Phi^{a} &=& \partial_{\mu} \Phi^{a} + \epsilon^{abc} A^{b}_{\mu}\Phi^{c},\\
\label{eq.2}
F^a_{\mu\nu} &=& \partial_{\mu}A^a_\nu - \partial_{\nu}A^a_\mu + \epsilon^{abc}A^b_{\mu}A^c_\nu.
\label{eq.3}
\end{eqnarray}
Since the gauge field coupling constant, can be scaled away, we set it to one without any loss of generality. The metric used is $g_{\mu\nu} = (- + + +)$. The SU(2) internal group indices $a, b, c$ run from 1 to 3 and the spatial indices are $\mu, \nu, \alpha = 0, 1, 2$, and $3$ in Minkowski space.

The equations of motion that follow from the Lagrangian (\ref{eq.1}) are
\begin{eqnarray}
D^{\mu}F^a_{\mu\nu} &=& \partial^{\mu}F^a_{\mu\nu} + \epsilon^{abc}A^{b\mu}F^c_{\mu\nu} = \epsilon^{abc}\Phi^{b}D_{\nu}\Phi^c,\nonumber\\
D^{\mu}D_{\mu}\Phi^a &=& \lambda\Phi^a(\Phi^{b}\Phi^{b} - \frac{\mu^2}{\lambda}).
\label{eq.4}
\end{eqnarray}
Non-BPS solutions to the YMH theory are obtained by solving the second order differential equations of motion (\ref{eq.4}), whereas BPS solutions can be more easily obtained by solving the Bogomol'nyi equations,
\begin{equation}
B^a_i \pm D_i\Phi^a = 0,
\label{eq.5}
\end{equation}
which is of first order.

The Abelian electromagnetic field tensor as proposed by 't Hooft \cite{kn:1} is
\begin{eqnarray}
F_{\mu\nu} &=& \hat{\Phi}^a F^a_{\mu\nu} - \epsilon^{abc}\hat{\Phi}^{a}D_{\mu}\hat{\Phi}^{b}D_{\nu}\hat{\Phi}^c\nonumber\\
&=& \partial_{\mu}A_\nu - \partial_{\nu}A_\mu - \epsilon^{abc}\hat{\Phi}^{a}\partial_{\mu}\hat{\Phi}^{b}\partial_{\nu}\hat{\Phi}^c,
\label{eq.6}
\end{eqnarray}
where $A_\mu = \hat{\Phi}^{a}A^a_\mu,~~\hat{\Phi}^a = \Phi^a/|\Phi|,~~|\Phi| = \sqrt{\Phi^{a}\Phi^{a}}$. Hence the 't Hooft electric field is $E_i = F_{i0}$, and the 't Hooft magnetic field is $B_i = -\frac{1}{2}\epsilon_{ijk}F_{jk}$, where the indices, $i, j, k = 1, 2, 3$. 
The topological magnetic current, which is also the topological current density of the system is \cite{kn:7} 
\begin{equation}
k_\mu = \frac{1}{8\pi}~\epsilon_{\mu\nu\rho\sigma}~\epsilon_{abc}~\partial^{\nu}\hat{\Phi}^{a}~\partial^{\rho}\hat{\Phi}^{b}~\partial^{\sigma}\hat{\Phi}^c.
\label{eq.7}
\end{equation}
Therefore the corresponding conserved topological magnetic charge is
\begin{eqnarray}
M = \int d^{3}x~k_0 
 =  \frac{1}{4\pi} \oint d^{2}\sigma_{i}~B_i. 
\label{eq.8}
\end{eqnarray}

In the BPS limit when the Higgs potential vanishes, the energy can be written in the form 
\begin{eqnarray}
E &=& \mp\int\partial_i(B^a_i\Phi^a)~d^3 x + \int\frac{1}{2}(B^a_i \pm D_i\Phi^a)^2~d^3 x\nonumber\\
&=& \mp\int\partial_i(B^a_i\Phi^a)~d^3 x = 4\pi M\frac{\mu}{\sqrt{\lambda}},
\label{eq.9}
\end{eqnarray}
where $M$ is the ``topological charge" when the vacuum expectation value of the Higgs field, $\frac{\mu}{\sqrt{\lambda}}$, is non zero coupled with some non-trivial topological structure of the fields at large $r$.

\section{The Dyons}
\label{sect:3}
\subsection{The Ansatz}
\label{sect:3.1}
The time independent gauge fields and Higgs field that leads to the MAP dyon solutions are given respectively by, \cite{kn:8}
\begin{eqnarray}
A_i^a &=&  - \frac{1}{r}\psi_1(r, \theta) \hat{n}^{a}_\phi\hat{\theta}_i + \frac{1}{r}\psi_2(r, \theta)\hat{n}^{a}_\theta\hat{\phi}_i
+ \frac{1}{r}R_1(r, \theta)\hat{n}^{a}_\phi\hat{r}_i - \frac{1}{r}R_2(r, \theta)\hat{n}^{a}_r\hat{\phi}_i, \nonumber\\
A^a_0 &=& {\cal A}_1(r, \theta)~\hat{n}^a_r + {\cal A}_2(r, \theta)\hat{n}^a_\theta,  \nonumber\\
\Phi^a &=& \Phi_1(r, \theta)~\hat{n}^a_r + \Phi_2(r, \theta)\hat{n}^a_\theta,
\label{eq.10}
\end{eqnarray}
\noindent where the spatial spherical coordinate orthonormal unit vectors are
\begin{eqnarray}
\hat{r}_i &=& \sin\theta ~\cos \phi ~\delta_{i1} + \sin\theta ~\sin \phi ~\delta_{i2} + \cos\theta~\delta_{i3}, \nonumber\\
\hat{\theta}_i &=& \cos\theta ~\cos \phi ~\delta_{i1} + \cos\theta ~\sin \phi ~\delta_{i2} - \sin\theta ~\delta_{i3}, \nonumber\\
\hat{\phi}_i &=& -\sin \phi ~\delta_{i1} + \cos \phi ~\delta_{i2},
\label{eq.11}
\end{eqnarray}
and the isospin coordinate orthonormal unit vectors are 
\begin{eqnarray}
\hat{n}_r^a &=& \sin \theta ~\cos n\phi ~\delta_{1}^a + \sin \theta ~\sin n\phi ~\delta_{2}^a + \cos \theta~\delta_{3}^a,\nonumber\\
\hat{n}_\theta^a &=& \cos \theta ~\cos n\phi ~\delta_{1}^a + \cos \theta ~\sin n\phi ~\delta_{2}^a - \sin \theta ~\delta_{3}^a,\nonumber\\
\hat{n}_\phi^a &=& -\sin n\phi ~\delta_{1}^a + \cos n\phi ~\delta_{2}^a; ~~~\mbox{where}~~n\geq 1.
\label{eq.12}
\end{eqnarray}
The constant parameter $-1\leq\eta\leq 1$ in Eq. (\ref{eq.10}) can also be written as $\eta=\tanh \gamma$. The $\phi$-winding number $n$ is a natural number. The MAP solutions exist only when $n=1, 2$. When $n=3$, the MAP disappeared and a vortex ring is formed instead \cite{kn:8}. Hence in our numerical calculation, we solved for the MAP dyon solutions only when $n=1, 2$.  

\subsection{The Solutions}
\label{sect:3.2}
In order to solve the for the MAP dyons solutions, the ansatz (\ref{eq.10}) is substituted into the equations of motion (\ref{eq.4}) and the 15 equations of motion are reduced to eight coupled second order partial differential equations. These eight equations are solved asymptotically first at small and then at large distances. In these asymptotic regions, the time component of the gauge field and the Higgs field are parallel in the isospin space, that is $\Phi_1 = {\cal A}_1$ and $\Phi_2 = {\cal A}_2$. The asymptotic solutions and boundary conditions at small distances that will give rise to finite energy MAP dyons solutions are
\begin{eqnarray}
\psi_A(0, \theta) = R_A(0, \theta) = 0, ~~~ A&=&1,2; \nonumber\\
\sin\theta ~\Phi_1(0,\theta) + \cos\theta ~\Phi_2(0,\theta) &=& 0\nonumber\\
\sin\theta ~{\cal A}_1(0,\theta) + \cos\theta ~{\cal A}_2(0,\theta) &=& 0\nonumber\\
\left.\frac{\partial}{\partial r}\left(\cos\theta ~\Phi_1(r,\theta) - \sin\theta ~\Phi_2(r,\theta)\right)\right|_{r=0} &=& 0 \nonumber\\
\left.\frac{\partial}{\partial r}\left(\cos\theta ~{\cal A}_1(r,\theta) - \sin\theta ~{\cal A}_2(r,\theta)\right)\right|_{r=0} &=& 0.
\label{eq.13}
\end{eqnarray}

At large distances, the asymptotic solution can be written as 

\begin{eqnarray}
\psi_1(\infty, \theta) = 2, ~~\psi_2(\infty, \theta) = 2n, ~~~
R_A(\infty, \theta) &=& 0, \nonumber\\
\Phi_1(\infty, \theta) = \xi\cos\theta, ~~~\Phi_2(\infty, \theta) &=& \xi\sin\theta, \nonumber\\
{\cal A}_1(\infty, \theta) = \eta \xi\cos\theta, ~~~{\cal A}_2(\infty, \theta) &=& \eta \xi\sin\theta
\label{eq.14}\\
&&\nonumber\\
R_A(r, \theta)|_{\theta \rightarrow 0, ~\pi} = \Phi_2(r, \theta)|_{\theta \rightarrow 0, ~\pi} = {\cal A}_2(r, \theta)|_{\theta \rightarrow 0, ~\pi} &=& 0, \nonumber\\
\partial_{\theta}\psi_A(r, \theta)|_{\theta \rightarrow 0, ~\pi} = \partial_{\theta}\Phi_1(r, \theta)|_{\theta \rightarrow 0, ~\pi} = \partial_{\theta}{\cal A}_1(r, \theta)|_{\theta \rightarrow 0, ~\pi} &=& 0,
\label{eq.15}
\end{eqnarray}
where $A=1, 2$, and the expectation value $\xi=\frac{\mu}{\sqrt{\lambda}}=1$.
The numerical MAP dyons solutions connecting the asymptotic solutions (\ref{eq.13}) to (\ref{eq.14}) and subjected to the boundary conditions (\ref{eq.15}) together with the gauge fixing condition \cite{kn:8}
\begin{equation}
\partial_rR_1-r\partial_\theta \psi_1=0,
\label{eq.16}
\end{equation}
were solved using the Maple 12 and MatLab R2009a softwares. The second order equations of motion (\ref{eq.4}) which are reduced to eight partial differential equations with the ansatz (\ref{eq.10}) are then transformed into a system of nonlinear equations using the finite difference approximation. This system of nonlinear equations are then discretized on a non-equidistant grid of size $70\times60$ covering the integration regions $0\leq \bar{x} \leq 1$ and $0\leq \theta \leq \pi$. Here $\bar{x}$ is the finite interval compactified coordinate given by $\bar{x}=\frac{r}{r+1}$. The partial derivative with respect to the radial coordinate is then replaced accordingly by ~$\partial_r \rightarrow (1-\bar{x})^2 \partial_{\bar{x}}$~ and ~$\frac{\partial^2}{\partial r^2} \rightarrow (1-\bar{x})^4\frac{\partial^2}{\partial \bar{x}^2} - 2(1-\bar{x})^3\frac{\partial}{\partial \bar{x}}$~. First of all, we used Maple to find the Jacobian sparsity pattern for the system of nonlinear equations.  After that we provide this information to Matlab to run the numerical computation. The system of nonlinear equations are then solved numerically using the trust-region-reflective algorithm by providing the solver with good initial guess.

The second order equations of motion Eq. (\ref{eq.4}) were solved when the $\phi$-winding number $n=1, 2$ and with zero Higgs potential, that is $\lambda=\mu=0$ but with nonzero expectation value $\xi=1$ and finally with $\lambda=\mu=1$ and expectation value $\xi=1$.

In the Prasad-Sommerfield limit where $\lambda=0$, the ansatz (\ref{eq.10}) can be simplified by letting 
\begin{eqnarray}
{\cal A}_1(r, \theta)=\eta\Phi_1(r, \theta) ~~\mbox{and}~~ {\cal A}_2(r, \theta)=\eta\Phi_2(r, \theta)
\label{eq.17}
\end{eqnarray}
hence making the gauge potential $A_0^a$ parallel to the Higgs field $\Phi^a$ in isospin space.

Upon substituting ansatz (\ref{eq.10}) and (\ref{eq.17}) into the equations of motion (\ref{eq.4}), the resulting equations of motion are simplified to,
\begin{eqnarray}
D^iF^a_{ij} = \partial^iF^a_{ij} + \epsilon^{abc}A^{bi}F^c_{ij} = \epsilon^{abc}\tilde{\Phi}^{b}D_j\tilde{\Phi}^c, ~~~D^iD_i\tilde{\Phi}^a = 0,
\label{eq.18}
\end{eqnarray}
where $\tilde{\Phi}^a = \sqrt{1-\eta^2} ~\Phi^a = \mbox{sech} \gamma ~\Phi^a$, the covariant time derivative $D_0\Phi^a = 0$,  and the electric field $E^a_i = F^a_{i0}= -\eta D_i\Phi^a$.
The time component of the 't Hooft gauge potential becomes,
$A_0 = \eta |\Phi|$.
Hence the 't Hooft electric field is non zero when $\eta\not=0$ and the monopole becomes a dyon. 

\subsection{The Magnetic Field, Magnetic Charge, and MAP Separation}
\label{sect:3.3}
\begin{table}[h]
\begin{center}
\begin{tabular}{|c|c|c|c|c|c|c|}
\hline
& \multicolumn{6}{|c|}{$\lambda = 0$}  \\ \cline{2-7}
& \multicolumn{3}{|c|}{n = 1} & \multicolumn{3}{|c|}{n = 2}  \\ \hline
$\eta$ & Q & d & E & Q & d & E \\ \hline
0 & 0 & 4.1970 & 1.6948 & 0 & 1.7944 & 2.9514  \\ \hline
0.20 & 4.3573 & 4.2840 & 1.7298 & 7.6013 & 1.8312 & 3.0126  \\ \hline
0.40 & 9.3156 & 4.5806 & 1.8496 & 16.2519 & 1.9588 & 3.2216  \\ \hline
0.60 & 16.0064 & 5.2502 & 2.1196 & 27.9272 & 2.2476 & 3.6928  \\ \hline
0.80 & 28.4471 & 7.0118 & 2.8268 & 49.6461 & 3.0084 & 4.9274  \\ \hline
0.90 & 44.0378 & 9.6884 & 3.8908 & 76.9414 & 4.1712 & 6.7909  \\ \hline
0.97 & 84.8568 & 17.6718 & 6.9571 & 150.4439 & 7.7512 & 12.3184  \\ \hline
0.98 & 105.2895 & 21.9170 & 8.5435 & 183.2625 & 9.6248 & 14.8613  \\ \hline

\end{tabular}

\end{center}
\caption{Values of $Q(n,\lambda,\eta)$, $d(n,\lambda,\eta)$, and $E(n,\lambda,\eta)$, when $\lambda=0$ and $n=1, 2$. When $\eta\rightarrow 1$, $Q$, $d$, and $E$ diverges.}
\label{table.1}
\end{table}

\begin{table}[h]
\begin{center}
\begin{tabular}{|c|c|c|c|c|c|c|}
\hline
& \multicolumn{6}{|c|}{$\lambda = 1$} \\ \cline{2-7}
& \multicolumn{3}{|c|}{n = 1} & \multicolumn{3}{|c|}{n = 2} \\ \hline
$\eta$ & Q & d & E & Q & d & E \\ \hline
0 & 0 & 3.2572 & 2.3635 & 0 & 1.4958 & 4.8582 \\ \hline
0.20 & 2.7387 & 3.2956 & 2.3854 & 3.7496 & 1.5122 & 4.8882 \\ \hline
0.40 & 5.6103 & 3.4204 & 2.4543 & 7.5969 & 1.5598 & 4.9805 \\ \hline
0.60 & 8.7866 & 3.6652 & 2.5813 & 11.6555 & 1.6518 & 5.1428 \\ \hline
0.80 & 12.5557 & 4.1264 & 2.7924 & 16.0816 & 1.8068 & 5.3906 \\ \hline
0.90 & 14.8390 & 4.5218 & 2.9470 & 18.5080 & 1.9210 & 5.5551 \\ \hline
0.97 & 16.7087 & 4.9368 & 3.0860 & 20.3277 & 2.0224 & 5.6908 \\ \hline
0.98 & 17.0015 & 5.0116 & 3.1087 & 20.5977 & 2.0384 & 5.7118 \\ \hline
1.00 & 17.6100 & 5.1760 & 3.1570 & 21.1500 & 2.0720 & 5.7550 \\ \hline

\end{tabular}

\end{center}
\caption{Values of $Q(n,\lambda,\eta)$, $d(n,\lambda,\eta)$, and $E(n,\lambda,\eta)$, when $\lambda=1$ and $n=1, 2$. When $\eta\rightarrow 1$, $Q$, $d$, and $E$ tends to their respective maximum critical values.}
\label{table.2}
\end{table}


\begin{figure}[tbh]
	\centering
		\hskip-0.4in \includegraphics[width=6.3in,height=4.6in]{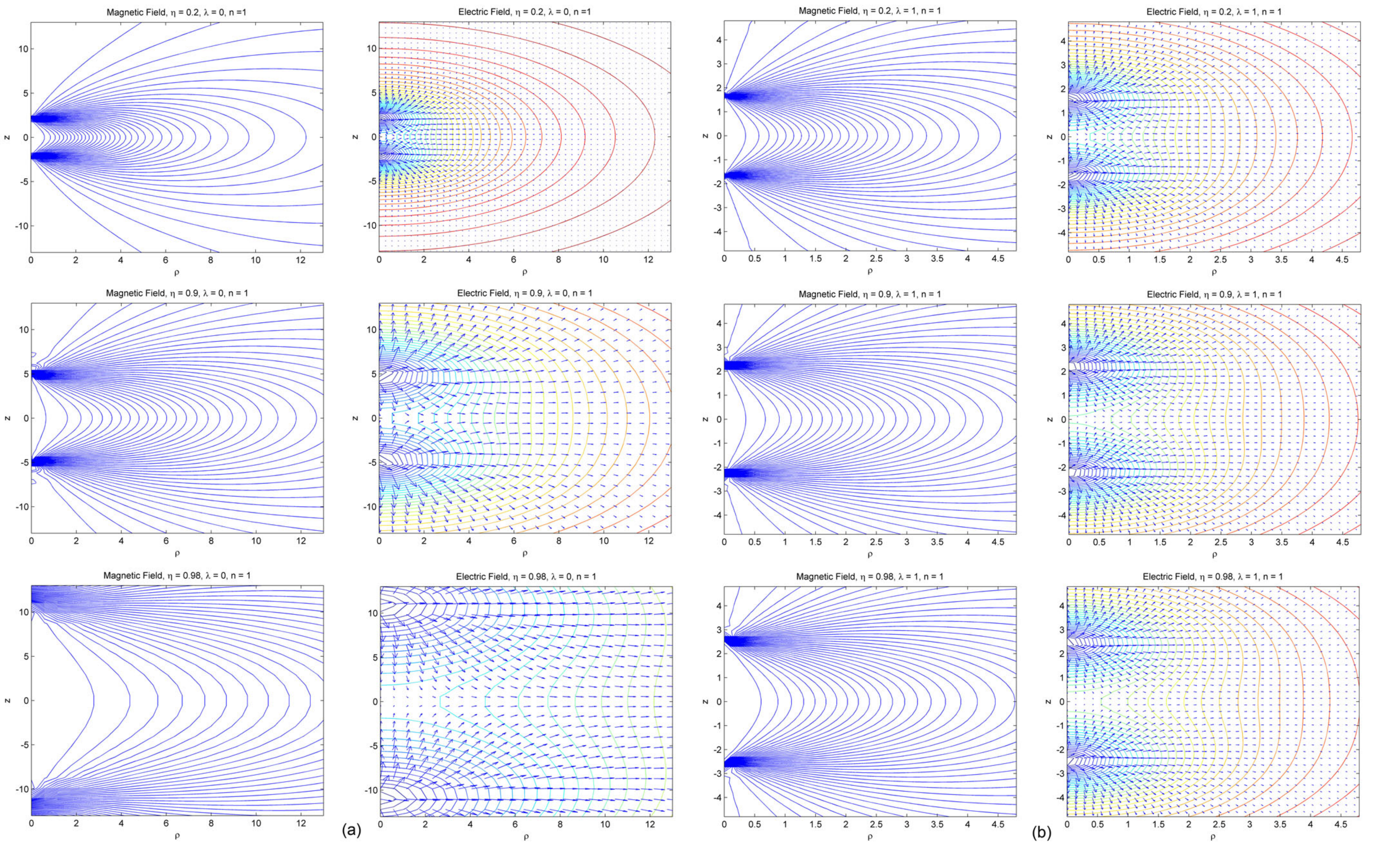}
	\caption{The magnetic field lines and the electric field plots of the MAP dyons when $n=1$ for (a) $\lambda=0$ and (b) $\lambda=1$ when $\eta=0.20, 0.90, 0.98$.}
	\label{fig.1}
\end{figure}


\begin{figure}[tbh]
	\centering
		\hskip-0.4in \includegraphics[width=6.3in,height=4.6in]{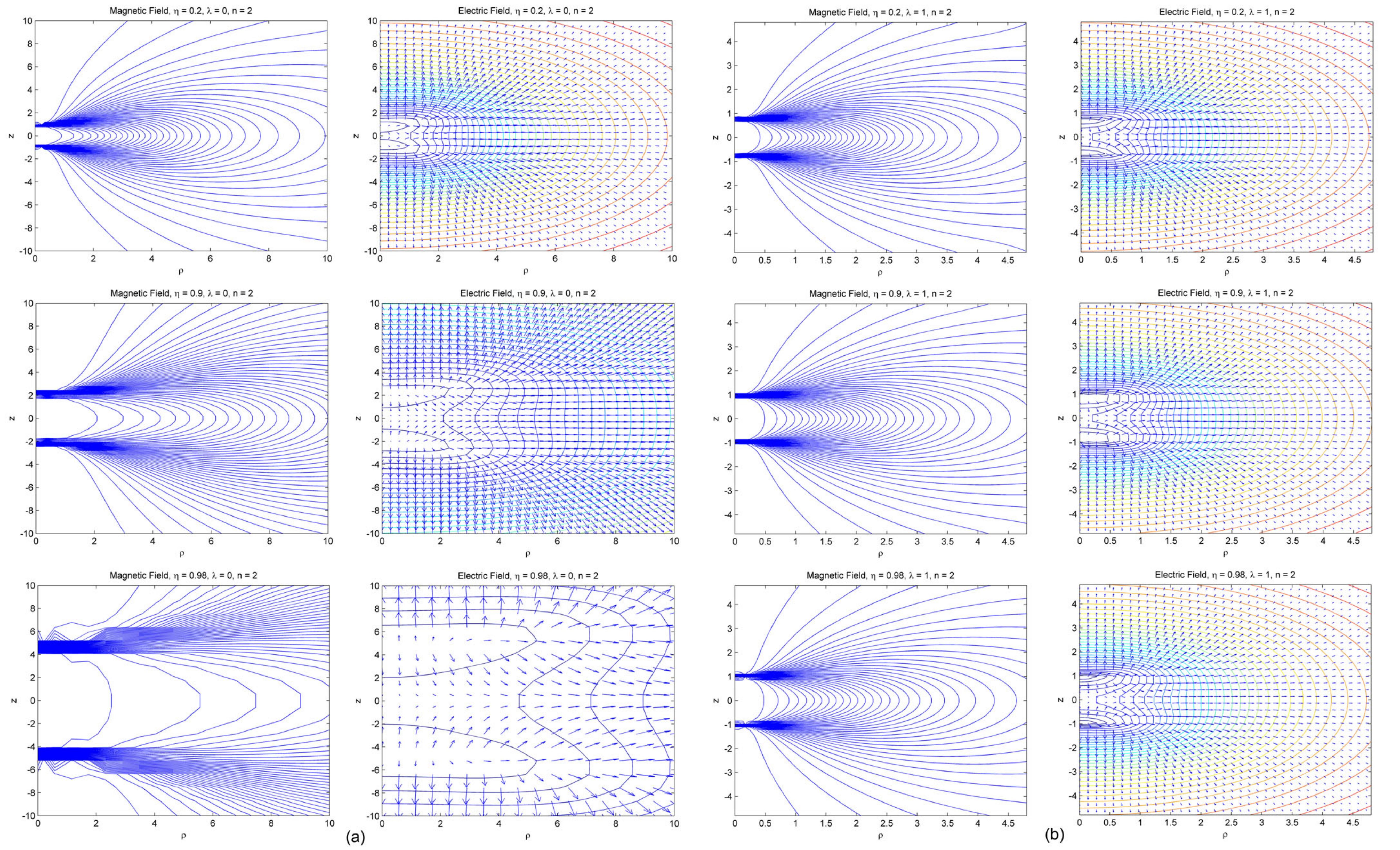}
	\caption{The magnetic field lines and the electric field plots of the MAP dyons when $n=2$ for (a) $\lambda=0$ and (b) $\lambda=1$ when $\eta=0.20, 0.90, 0.98$.}
	\label{fig.2}
\end{figure}

The 't Hooft magnetic field $B_i$ is made up of two parts, the gauge part and the Higgs part, 
\begin{eqnarray}
B_i &=& B^G_i + B^H_i, \nonumber\\
B^G_i &=& -\epsilon_{ijk}\partial_jA_k  ~~~\mbox{and}~~~B^H_i = \frac{1}{2}\epsilon_{ijk}\epsilon^{abc}\hat{\Phi}^{a}\partial_{j}\hat{\Phi}^{b}\partial_{k}\hat{\Phi}^c.
\label{eq.19}
\end{eqnarray}
With some calculations, the gauge part of the magnetic field is given by
\begin{eqnarray}
B^G_i &=& -n\epsilon_{ijk}\partial_j\sin\kappa ~\partial_k \phi, ~~\nonumber\\
\mbox{where},~~\sin\kappa &=& \frac{\sin\theta}{n}\left(\psi_2 \frac{\Phi_2}{|\Phi|} -R_2 \frac{\Phi_1}{|\Phi|}\right),
\label{eq.20}
\end{eqnarray}
To calculate for the 't Hooft magnetic field $B_i^H$, we rewrite the Higgs field from the spherical to the Cartesian coordinate system, 
\begin{eqnarray}
\Phi^a &=& \Phi_{1}~\hat{n}^{a}_r + \Phi_{2}~\hat{n}^{a}_{\theta} + \Phi_3~\hat{n}^{a}_{\phi}\nonumber\\
&=& \tilde{\Phi}_1 ~\delta^{a1} + \tilde{\Phi}_2 ~\delta^{a2} + \tilde{\Phi}_3 ~\delta^{a3},
\label{eq.21}
\end{eqnarray}
\begin{eqnarray}
\mbox{where}~~~\tilde{\Phi}_1 &=& \sin\theta \cos n\phi ~\Phi_1 + \cos\theta \cos n\phi ~\Phi_2 - \sin n\phi ~\Phi_3
= |\Phi|\cos\alpha \sin\beta\nonumber\\
\tilde{\Phi}_2 &=& \sin\theta \sin n\phi ~\Phi_1 + \cos\theta \sin n\phi ~\Phi_2 + \cos n\phi ~\Phi_3
= |\Phi|\cos\alpha \cos\beta\nonumber\\
\tilde{\Phi}_3 &=& \cos\theta ~\Phi_1 - \sin\theta ~\Phi_2 = |\Phi|\sin\alpha.
\label{eq.22}
\end{eqnarray}
The Higgs unit vector can be simplified to 
\begin{eqnarray}
\hat{\Phi}^a &=& \cos\alpha \sin\beta ~\delta^{a1} + \cos\alpha \cos\beta ~\delta^{a2} + \sin\alpha ~\delta^{a3},\nonumber\\
\sin\alpha &=& \frac{\Phi_1}{|\Phi|}\cos\theta - \frac{\Phi_2}{|\Phi|}\sin\theta,~~~\beta=\frac{\pi}{2}-n\phi,
\label{eq.23}
\end{eqnarray}
and the 't Hooft magnetic field is reduced to only the $\hat{r}_i$ and $\hat{\theta}_i$ components,
\begin{eqnarray}
B_i^H &=& -n\epsilon_{ijk} \partial^j\sin\alpha\partial^k\phi \nonumber\\
&=& -\frac{n}{r^2 \sin\theta}\left\{\frac{\partial\sin\alpha}{\partial\theta}\right\}\hat{r}_i + 
\frac{n}{r\sin\theta}\left\{\frac{\partial\sin\alpha}{\partial r}\right\}\hat{\theta}_i.
\label{eq.24}
\end{eqnarray}
From our numerical calulation, we managed to plot magnetic field lines of the MAP dyons when $n=1$ for values of $\lambda=0, 1$ and $\eta=0.20, 0.90, 0.98$ as in Figure \ref{fig.1} and when $n=2$ for values of $\lambda=0, 1$ and $\eta=0.20, 0.90, 0.98$ as in Figure \ref{fig.2}.

From Eq.(\ref{eq.8}), the net magnetic charges enclosed by the upper and lower hemi-sphere at infinity are calculated to be, 
\begin{eqnarray}
M_+ &=& -\left.\frac{n}{2}\sin\alpha\right|^\frac{\pi}{2}_{0, r\rightarrow \infty} = +n,  \nonumber\\
M_- &=& -\left.\frac{n}{2}\sin\alpha\right|^\pi_{\frac{\pi}{2}, r\rightarrow \infty} = -n, ~~n=1,2,
\label{eq.25}
\end{eqnarray}
respectively. Hence the magnetic charge in the upper and lower hemi-sphere is $+n$ and $-n$ respectively and the net magnetic charge of the MAP dyon is zero. For the MAP dyon to exist, $n=1, 2$ and when $n=3$, the MAP dyon cease to exist and a vortex ring dyon is formed instead \cite{kn:8}. Following 't Hooft's definition for the electromagnetic field, Eq. (\ref{eq.6}), the magnetic charges of $n=\pm 1, \pm 2$, are all located at the zeros of the Higgs field and the MAP separations $d(n, \lambda, \eta)$ for different values of $n$, $\lambda$, and $\eta$ are as given in Table \ref{table.1} and \ref{table.2}.

From Figure \ref{fig.1} and Figure \ref{fig.2}, we can generally conclude that when $\lambda=1$, the MAP separations are smaller than when $\lambda=0$ for given values of $\eta$ and $n$ as the interaction between Higgs field is attractive. When $n=2$, the MAP separations are smaller than when $n=1$ for fixed values of $\lambda$ and $\eta$ as the MAP possess unlike magnetic charges that attract. When the value of $\eta$ increases the electric charges also increases and for fixed values of $\lambda$ and $n$, the MAP separations get wider as the MAP possess similiar electric charges that repel. Hence the MAP separation $d(n, \lambda, \eta)$ varies with $n$, $\lambda$, and $\eta$. From our numerical calculations, we found that when $\lambda=0$, $d(n, 0, \eta)$ increases with $\eta$ from $d(1, 0, 0)= 4.20$ and $d(2, 0, 0)= 1.79$ to infinity exponentially fast as $\eta$ approaches one. However when $\lambda=1$, $d(n, 1, \eta)$ increases with $\eta$ from $d(1, 1, 0)=3.26 $ and $d(2, 1, 0)= 1.50$ to finite critical values of $d(1, 1, 1)= 5.18$ and $d(2, 1, 1)= 2.07$ respectively as $\eta$ approaches one. See Table \ref{table.1} and \ref{table.2}. 

\subsection{The Electric Field and Electric Charge}
\label{sect:3.4}


\begin{figure}[tbh]
	\centering
	 \includegraphics[width=6in,height=5.5in]{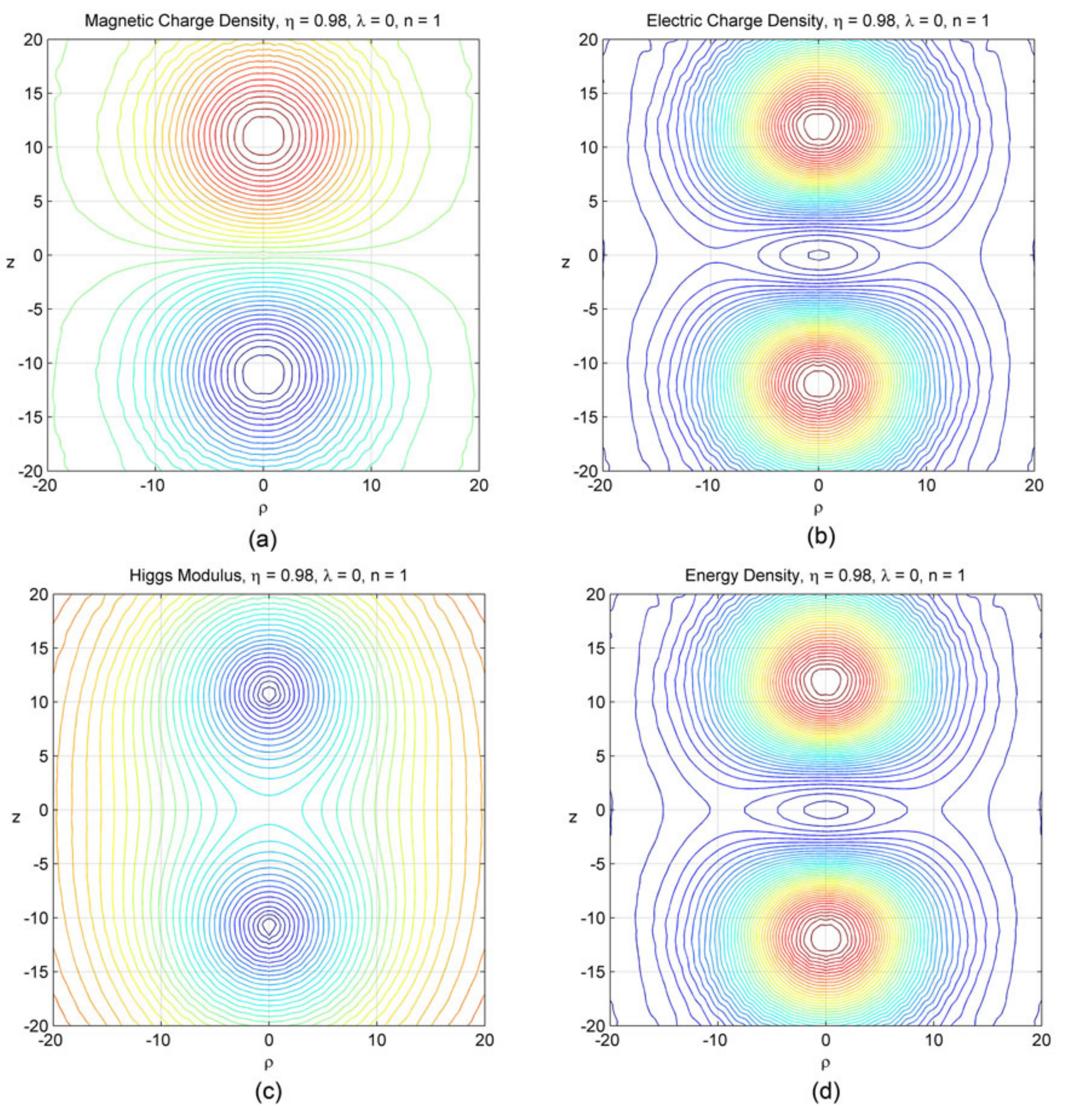} 
	\caption{Contour plots of (a) the magnetic charge density, (b) the electric charge density, (c) the Higgs field modulus, and (d) the energy density, when $n=1$ for $\lambda=0$ and $\eta =0.98$.}
	\label{fig.3}
\end{figure}


\begin{figure}[tbh]
	\centering
	 \includegraphics[width=6in,height=5.5in]{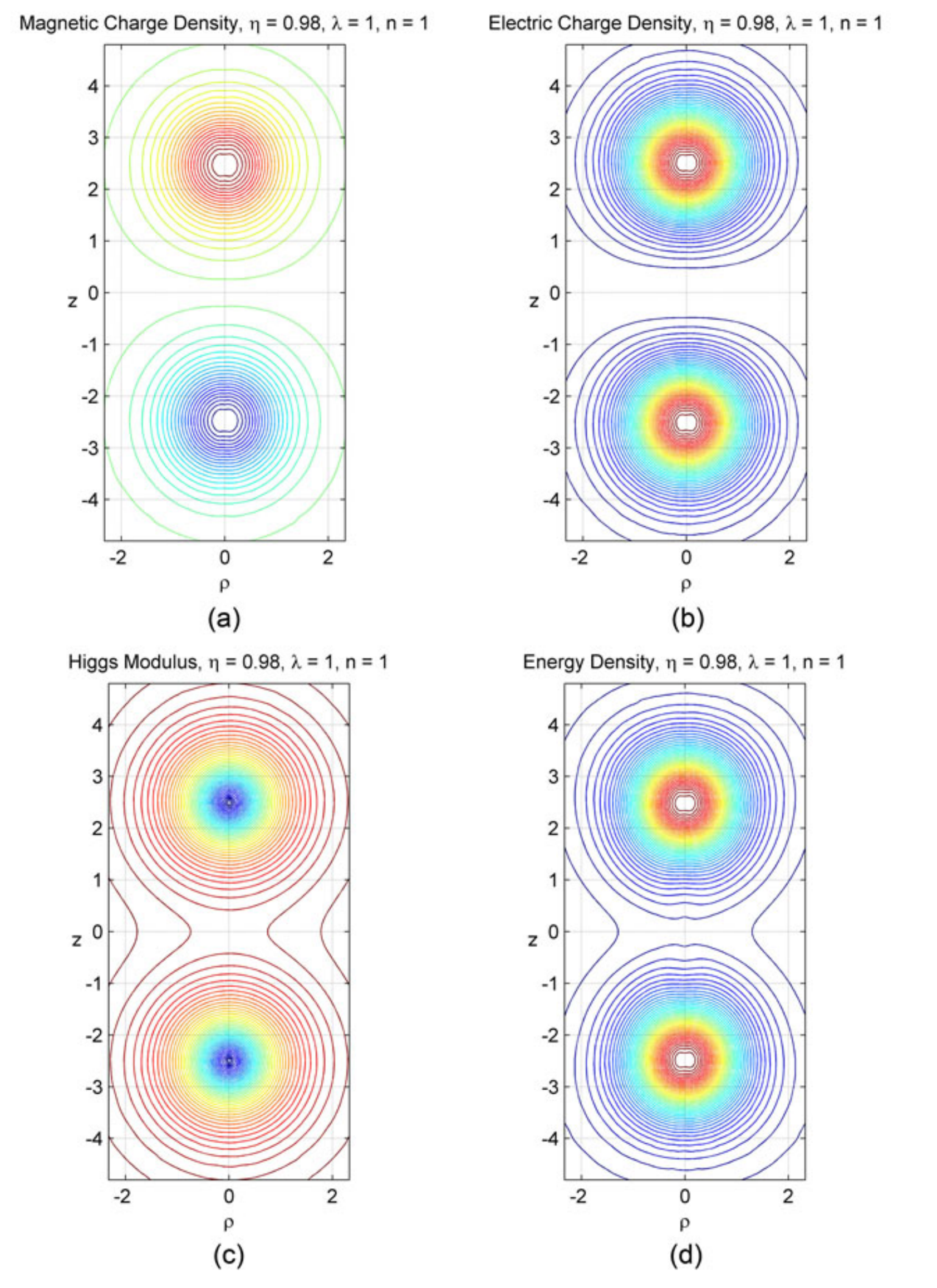} 
	\caption{Contour plots of (a) the magnetic charge density, (b) the electric charge density, (c) the Higgs field modulus, and (d) the energy density, when $n=1$ for $\lambda=1$ and $\eta =0.98$.}
	\label{fig.4}
\end{figure}


\begin{figure}[tbh]
	\centering
	 \includegraphics[width=6in,height=5.5in]{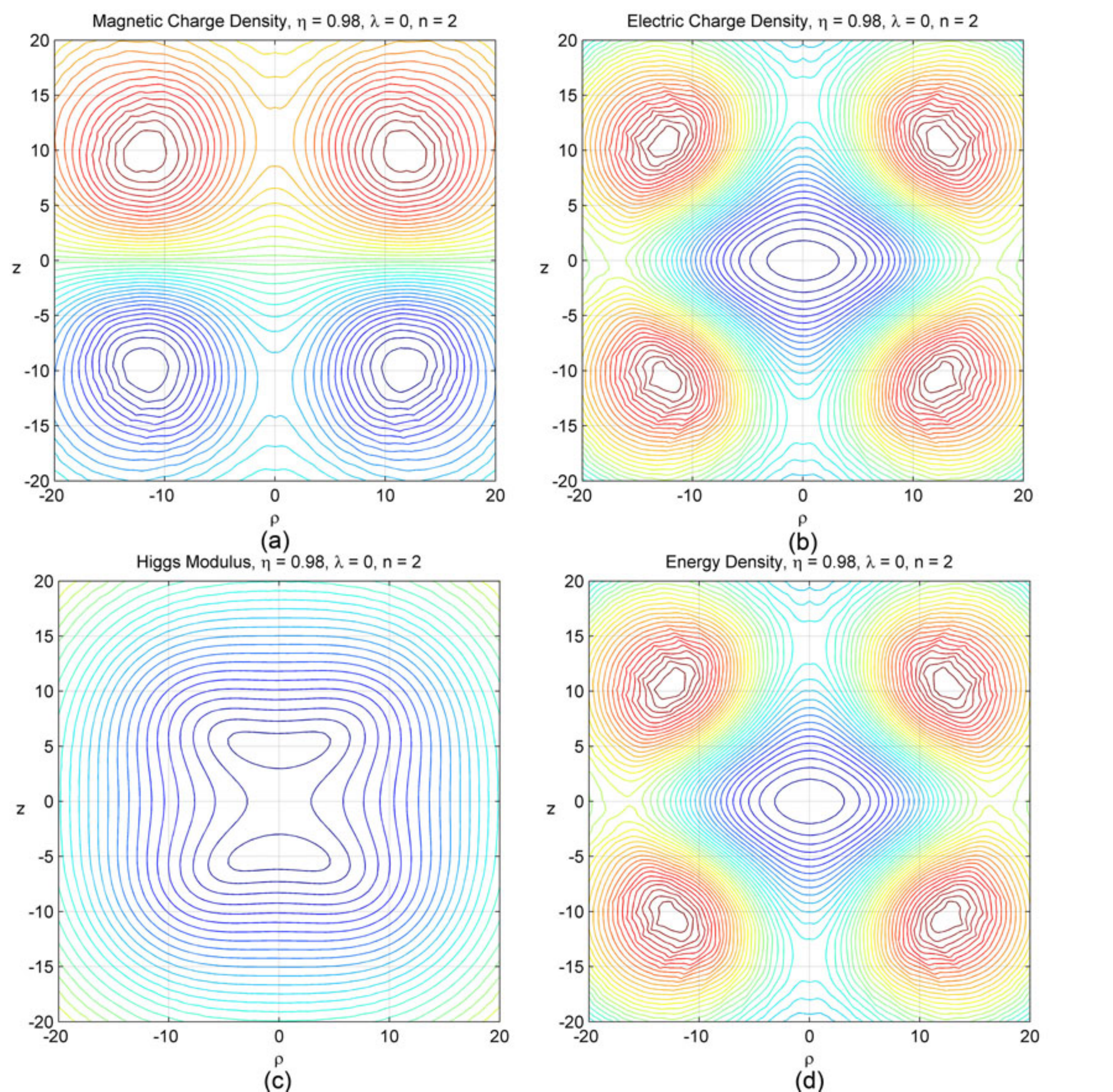} 
	\caption{Contour plots of (a) the magnetic charge density, (b) the electric charge density, (c) the Higgs field modulus, and (d) the energy density, when $n=2$ for $\lambda=0$ and $\eta =0.98$.}
	\label{fig.5}
\end{figure}


\begin{figure}[tbh]
	\centering
	 \includegraphics[width=6in,height=5.5in]{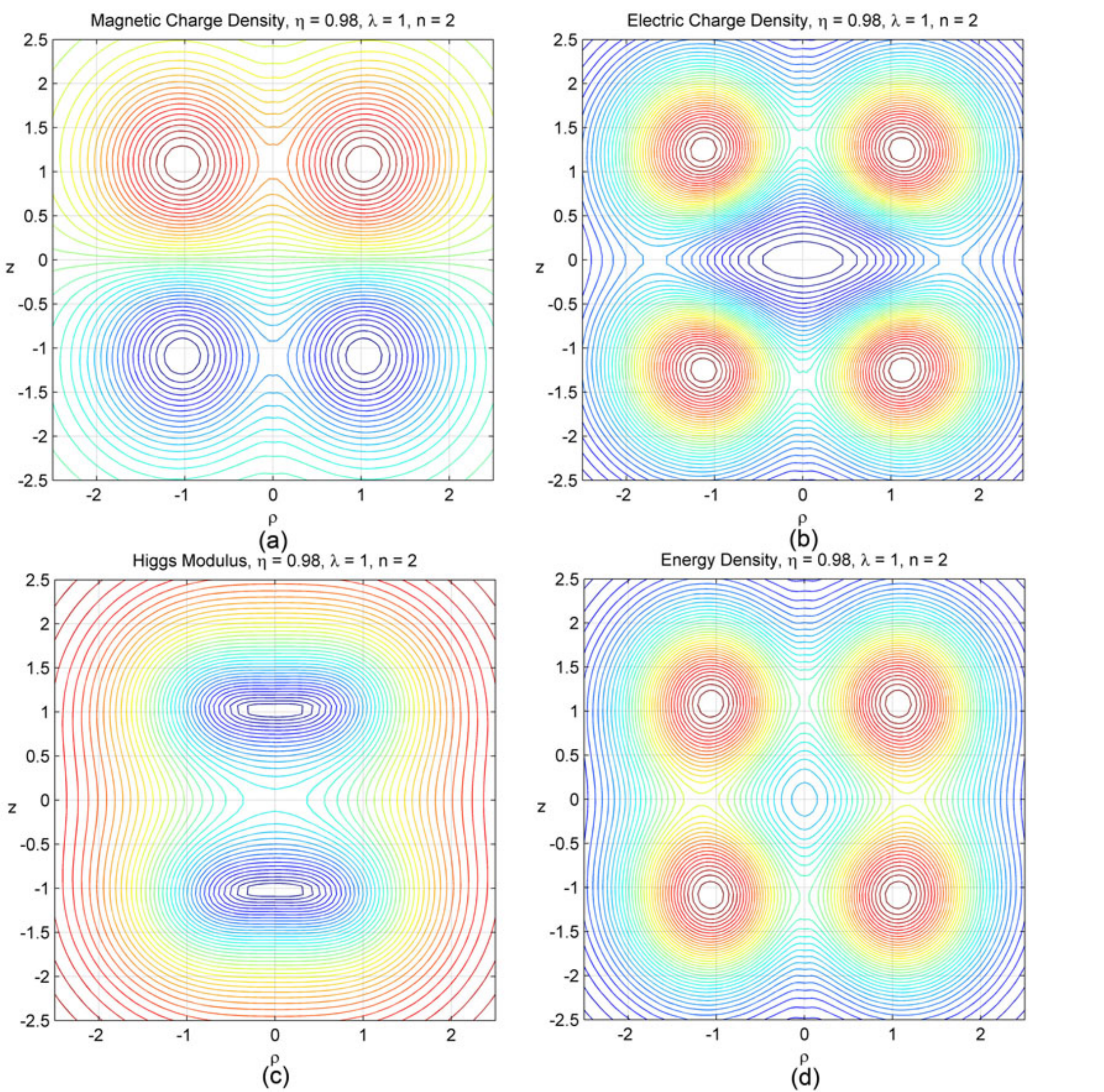} 
	\caption{Contour plots of (a) the magnetic charge density, (b) the electric charge density, (c) the Higgs field modulus, and (d) the energy density, when $n=2$ for $\lambda=1$ and $\eta =0.98$.}
	\label{fig.6}
\end{figure}


\begin{figure}[tbh]
	\centering
	\hskip-0.2in	 \includegraphics[width=6in,height=5.5in]{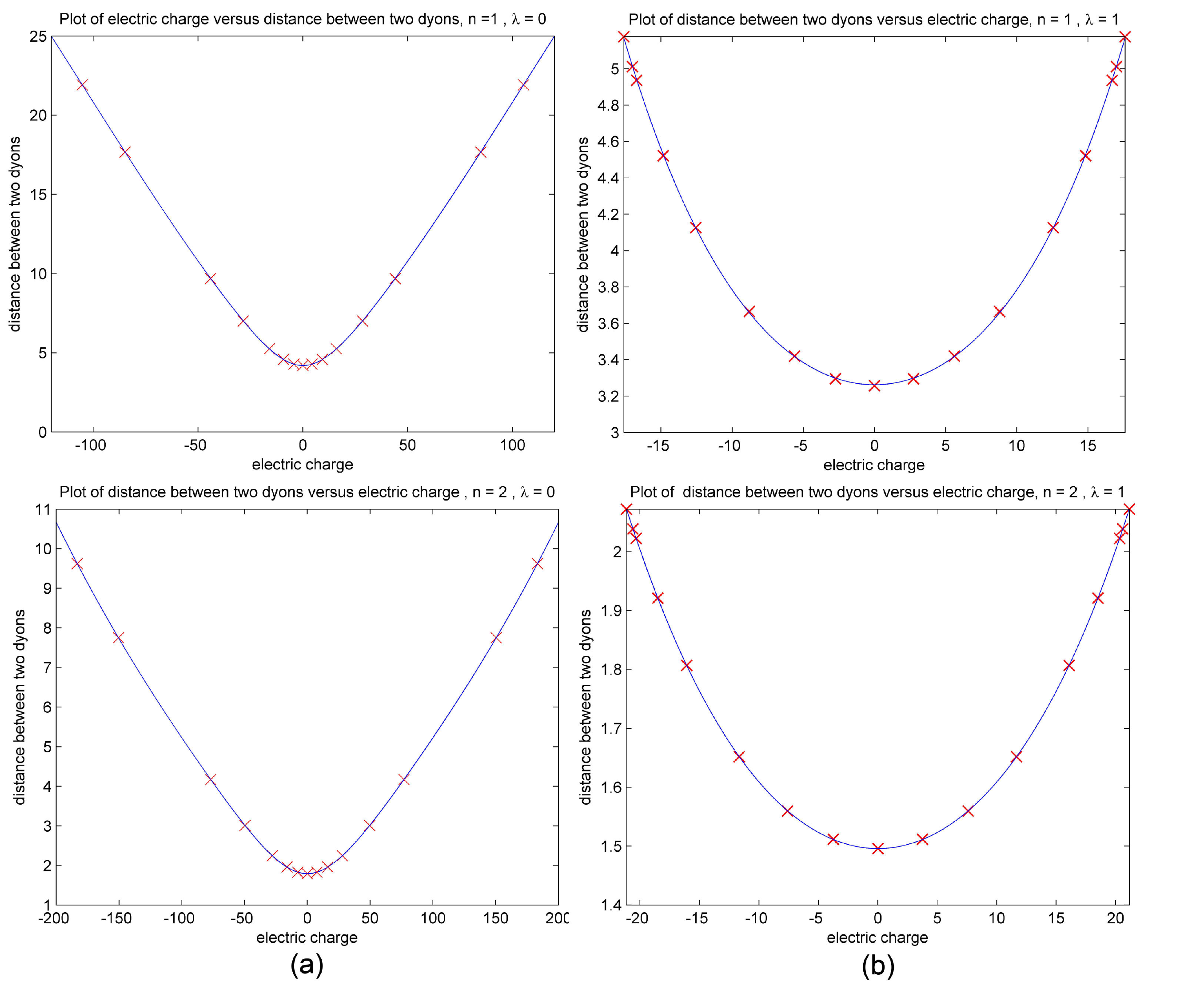} 
	\caption{Parametric plots of (a) $d(n,0,\eta)$ versus $Q(n,0,\eta)$ and (b) $d(n,1,\eta)$ versus $Q(n,1,\eta)$ when $n=1, 2$ for $-1<\eta<1$.}
	\label{fig.7}
\end{figure}

At spatial infinity in the Higgs vacuum, the electromagnetic field can be represented uniquely by 
\begin{eqnarray}
{\cal F}_{\mu\nu} &=& F_{\mu\nu}^a\left\{\frac{\Phi^a}{\xi}\right\},
\label{eq.26}
\end{eqnarray}
which corresponds to the unbroken U(1) symmetry. The corresponding electric and magnetic fields are then ${\cal E}_i = {\cal F}_{i0}$ and ${\cal B}_i = -\frac{1}{2}\epsilon_{ijk}{\cal F}_{jk}$. However according to Coleman \cite{kn:9}, there is no unique way of representing the electromagnetic field in the region of the monopole outside the Higgs vacuum. One proposal was as given by 't Hooft as in Eq. (\ref{eq.6}) which will reduced to Eq. (\ref{eq.26}) at spatial infinity in the Higgs vacuum. The 't Hooft's magnetic field has the special property, $\partial^i B_i=0$ when $|\Phi|\not=0$. Hence with 't Hooft's definition of the electromagnetic field, the magnetic charges can only reside at the zeros of the Higgs field and the magnetic charges of the MAP dyon configurations are concentrated at two points along the $z$-axis where the Higgs field vanishes. Thus there are points singularities in 't Hooft's definition of Eq. (\ref{eq.6}), which makes it not possible to calculate and plot the magnetic and electric charge density distributions numerically with this definition.

The 't Hooft electric field of the MAP dyon,  
$E_i = \partial_i A_0 = \partial_i\left\{{\cal A}_1\frac{\Phi_1}{|\Phi|} + {\cal A}_2\frac{\Phi_2}{|\Phi|}\right\}$ corresponds to Eq. (\ref{eq.26}) in the Higgs vacuum. 2D vector plots of the electric field $E_i$ are given in Figure \ref{fig.1} when $n=1$ and Figure \ref{fig.2} when $n=2$, for $\eta=0.20, 0.90, 0.98$ and (a) $\lambda=0$, (b) $\lambda=1$.
Unlike the magnetic field, the electric field varies proportionally with the constant $-1\leq\eta\leq 1$. Hence the electric field can be switched off by setting $\eta=0$. As the parameter $\eta$
increases from $-1$ to zero, the electric field points radially inwards and when $0<\eta\leq 1$, the electric field is radially outwards. Hence the electric charges of the both the monopole and antimonopole are positive (negative) when $\eta$ is positive (negative).

Another proposal for the electromagnetic field, ${\cal F}_{\mu\nu} = F_{\mu\nu}^a\left\{\frac{\Phi^a}{\eta}\right\}$, was given by Bogolmolny \cite{kn:10} and Faddeev \cite{kn:11}. Using this definition, Eq. (\ref{eq.26}), the magnetic charge density $g=\partial^i {\cal B}_i$ and electric charge density $q=\partial^i {\cal E}_i$ distributions can be calculated and plotted numerically as shown in Figure \ref{fig.3} (a), (b) to Figure \ref{fig.6} (a), (b) when $\lambda=0, 1$ and $n=1, 2$ respectively.
The electric charge density, $q=\partial^i {\cal E}_i$, of the MAP dyon solutions is solely positive (negative) throughout space when $\eta$ is positive (negative).  When $n=1$, the electric charge distribution of the MAP dyon is two spheres centered at the two point zeros of the Higgs field. Hence the soliton has a dumbbell shape structure.

However when $n=2$, the electric charge distribution is two horizontal toruses around the two zeros of the Higgs field as center with the $z$-axis as the symmetry axis. Contour plots of the electric charge density distribution when $n=1, 2$,  $\lambda=0, 1$, and $\eta=0.98$ are given in Figure \ref{fig.3} (b) to Figure \ref{fig.6} (b). Contour plots of the magnetic charge density distribution, the Higgs field modulus, and energy density when $n=1, 2$, $\lambda=0, 1$, and $\eta=0.98$ are also given in Figure \ref{fig.3} (a), (c), (d) to Figure \ref{fig.6} (a), (c), (d) respectively. 

From Gauss' law, the total electric charge of the dyon is given by,
\begin{eqnarray}
Q(n, \lambda, \eta) = \int_{r\rightarrow \infty}{\cal E}_i\hat{r}_i~r^2\sin\theta~d\theta d\phi.
\label{eq.27}
\end{eqnarray}
The different values of $Q(n, \lambda, \eta)$ calculated for various values of $n$, $\lambda$, and $\eta$ are as tabulated in Table \ref{table.1} and \ref{table.2}. For fixed value of $n=1, 2$ and $\lambda=0$, $Q(\eta)$ increases with $\eta$ from $Q(n, 0, 0) = 0$ and approaches infinity exponentially fast as $\eta$ approaches one. However for fixed value of $n=1, 2$ and $\lambda=1$, $Q(\eta)$ increases with $\eta$ from $Q(n, 1, 0) = 0$ to a maximum critical value of $Q(1,1,\eta)=17.61$ and $Q(2,1,\eta)=21.15$ as $\eta$ approaches one. The four parametric graphs for MAP separation $d(n,\lambda, \eta)$ versus total electric charge $Q(n,\lambda, \eta)$ for $n=1, 2$ and $\lambda=0, 1$ were plotted for $-0.98<\eta<0.98$ when $\lambda=0$, Figure \ref{fig.7} (a) and for $-1<\eta< 1$ when $\lambda=1$, Figure \ref{fig.7} (b). The end points for the graphs when $\eta \rightarrow \pm 1$ in Figure \ref{fig.7} (b) are $(5.18, \pm 17.61)$ for $n=1$ and $(2.07, \pm 21.15)$ for $n=2$ whereas there is no end point for the graphs in Figure \ref{fig.7} (a) as $d(n,\lambda, \eta)$ and $Q(n,\lambda, \eta)$ diverge as $\eta \rightarrow \pm 1$. The relationship between the MAP separation $d(n,\lambda, \eta)$ and the total charge $Q(n,\lambda, \eta)$ when $\lambda=1$ is given by 
\begin{eqnarray}
d &\approx& 4.4360\times10^{-6}Q^4-9.2560\times10^{-20}Q^3+0.4771\times10^{-2}Q^2 \nonumber\\
  &+&2.7050\times10^{-18}Q+3.2630, ~~n=1, \nonumber\\
d &\approx& 4.5710\times10^{-7}Q^4+6.6840\times10^{-20}Q^3+0.1084\times10^{-2}Q^2 \nonumber\\
  &-&2.7650\times10^{-17}Q+1.4960, ~~n=2.
\label{eq.28}
\end{eqnarray}

\subsection{The Energy}
\label{sect:3.5}


\begin{figure}[tbh]
	\centering
	\hskip-0.2in	\includegraphics[width=5.8in,height=4.6in]{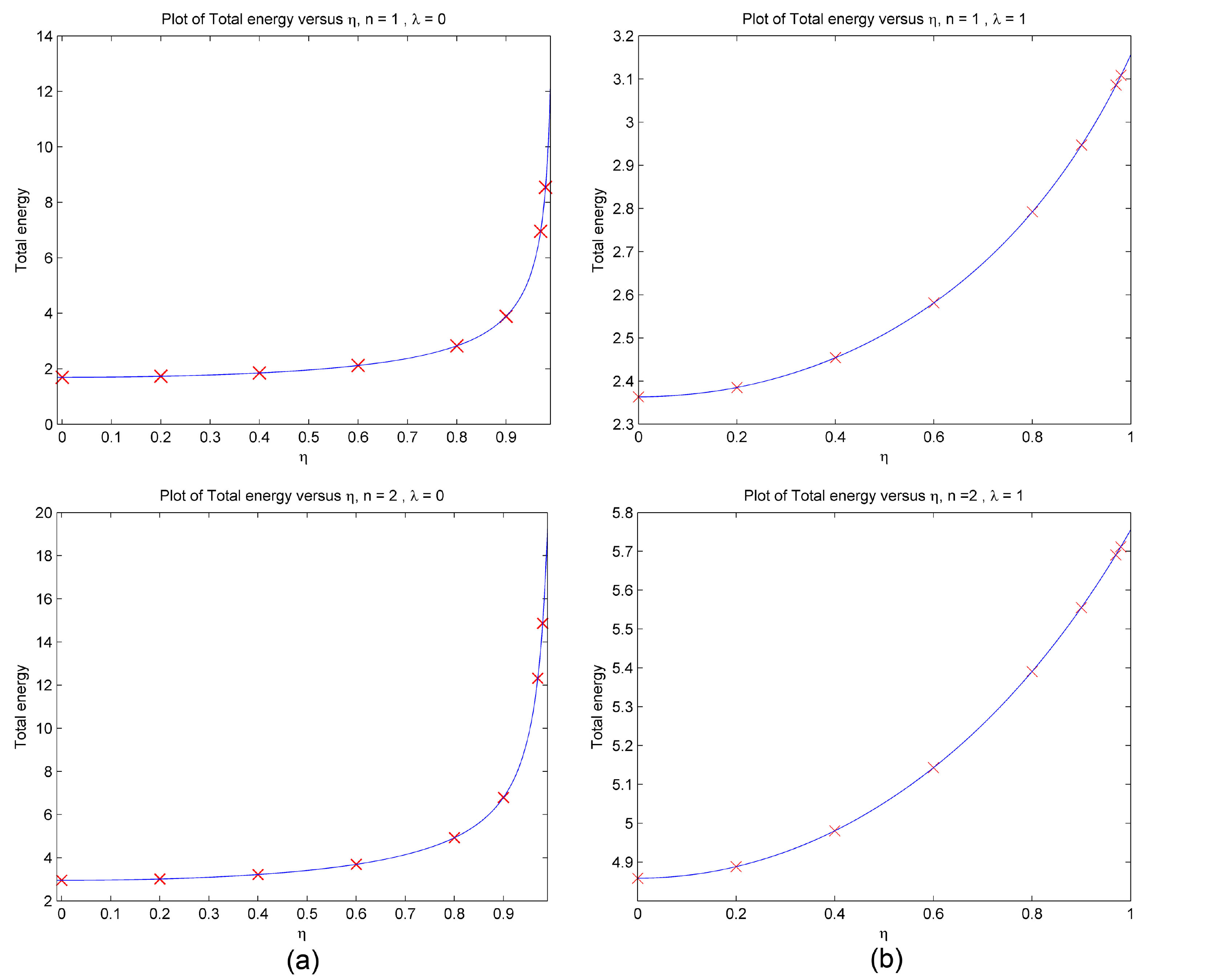}
	\caption{Plots of total energy $E$ versus $\eta$ when (a) $\lambda=0$ and (b) $\lambda=1$ for $n=1, 2$.}
	\label{fig.8}
\end{figure}


\begin{figure}[tbh]
	\centering
	\hskip-0.2in	\includegraphics[width=6.3in,height=3.0in]{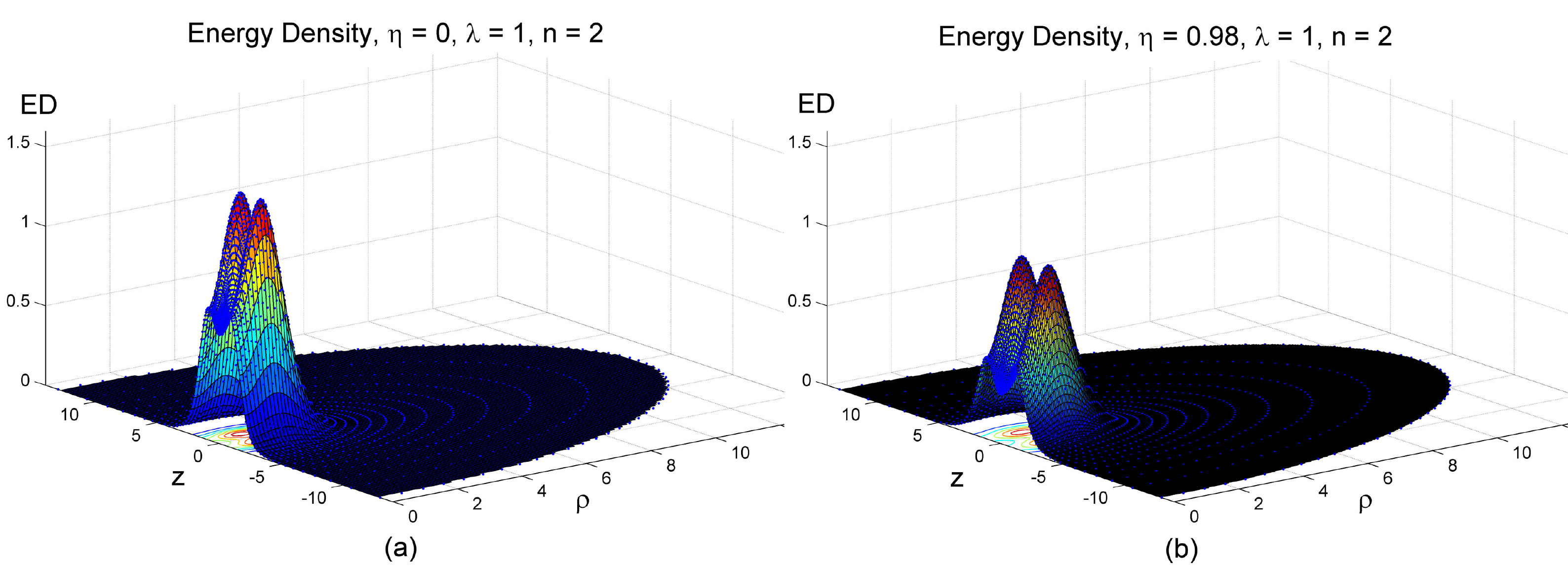}
	\caption{3D plots of the energy density distribution  for $n=2$, $\lambda=1$ when (a) $\eta=0$ and  (b) $\eta=0.98$.}
	\label{fig.9}
\end{figure}

Since the MAP dyon is not a BPS solution, its energy cannot be given by Eq. (\ref{eq.9}) but is given by
\begin{eqnarray}
E=\frac{1}{2}\int \{B^a_iB^a_i + E^a_iE^a_i + D_i\Phi^aD_i\Phi^a + D_0\Phi^aD_0\Phi^a + \frac{\lambda}{2}(\Phi^a\Phi^a-\xi^2)^2\}~d^3x.
\label{eq.29}
\end{eqnarray}
The different values of $E(n, \lambda, \eta)$ calculated when $\xi=1$ for different values of $n$, $\lambda$, and $\eta$ are as tabulated in Table \ref{table.1} and \ref{table.2}. Similar to the MAP separations $d$ and total electric charge $Q$, the total energy $E$ approaches a finite critical value when $\eta \rightarrow \pm 1$ for $\lambda=1$ as shown in Figure \ref{fig.8} (b). When $\lambda=0$, the total energy $E$ diverges exponentially fast as $\eta \rightarrow \pm 1$. The critical values of the total energy are $E(1, 1, \pm 1)=3.157$ and $E(2, 1, \pm 1)=5.755$, see Figure \ref{fig.8} (b). The energy density distribution is plotted 3D for $n=2$, $\lambda=1$ in Figure \ref{fig.7} (a) when $\eta=0$ and (b) when $\eta=0.98$. These plots show that the energy density of the MAP has a toroidal distribution around the two zeros of the Higgs field.

\section{Comments}
\label{sect.4}

The ansatz (\ref{eq.10}) used to construct the dyon solutions is introduced in the same standard way as in the construction of the Julia-Zee dyons years ago \cite{kn:3}. Hence when the parameter $\eta$ is zero, the electric field is switched off and only the magnetic field remains and the dyon becomes a monopole.

The MAP separation $d$, total electric charge $Q$, and the total energy $E$ of these MAP dyon solutions are infinite as $\eta\rightarrow \pm 1$ when $\lambda=0$ or when the Higgs field $\Phi^a$ is parallel to the time component of the gauge field $A^a_0$ in isospin space. However when $\Phi^a$ is not parallel to $A^a_0$, that is when $\lambda=1$, all the three quantities, $d(n,\lambda, \eta)$, $Q(n,\lambda, \eta)$, and $E(n,\lambda, \eta)$ approach critical values when $\eta\rightarrow \pm 1$ . The relationship between $d$, $Q$, and $E$ are been studied and $\{d(n,1,\eta)-d(n,1,0)\}$ is approximately proportional to $Q^2(n,1,\eta)$.

These MAP dyon solutions are axially symmetric about the $z$-axis. Hence the MAP dyon is a magnetic dipole with both poles carrying the same electric charges but opposite magnetic charges. When $n=1$, the electric and magnetic charges are two spheres with zeros of the Higgs field as center. The sphere in the upper hemishphere possesses positive magnetic and electric charges whereas the sphere in the lower hemisphere possesses negative magnetic charge and positive electric charge when $\eta$ is positive, Figure \ref{fig.3} and Figure \ref{fig.4}. The MAP dyons possess dumbbell shape mass distribution. When $\eta$ is negative, only the electric charges will reverse its sign. However when $n=2$, the magnetic and electric charges are two toruses with the zeros of the Higgs field as their center. In this case, the two zeros of the Higgs field are multiple zeros as shown in Figure \ref{fig.5} (c) and Figure \ref{fig.6} (c). Hence the soliton has the unique energy density distribution as given in Figure \ref{fig.5} (d) and Figure \ref{fig.6} (d) when $\lambda=0$ and $\lambda=1$ respectively. 

Our conclusion is that the MAP dyon when $n=2$ is not a two points monopole-antimonopole but rather, it is composed of two toruses of the same electric charges but opposite magnetic charges, whereas when $n=1$, the solition is composed of two spheres.

\section{Acknowlegements}
The authors would like to thank the Ministry of Science, Technology and Innovation (MOSTI) of Malaysia for the award of ScienceFund research grant (Project Number: 06-01-05-SF0266).

\end{document}